\documentclass{PoS}
\usepackage[numbers]{natbib}
\usepackage{slashed}
\DeclareGraphicsExtensions{.png,.pdf,.gif,.jpg}
\newcommand\Tr{\mathop{\rm Tr}}
\newcommand\Dslash{\slashed D}
\def\firstword#1 #2\endpar{#1}

\title{Quark Chromoelectric Dipole Moment Contribution to the
       Neutron Electric Dipole Moment}

\ShortTitle{qCEDM contribution to nEDM}

\author{\speaker{Tanmoy Bhattacharya}\\%\thanks{A footnote may follow.}\\
        Los Alamos National Laboratory\\
        E-mail: \email{tanmoy@lanl.gov}}

\author{Vincenzo Cirigliano\\
        Los Alamos National Laboratory\\
        E-mail: \email{cirigliano@lanl.gov}}
\author{Rajan Gupta\\
        Los Alamos National Laboratory\\
        E-mail: \email{rajan@lanl.gov}}
\author{Boram Yoon\\
        Los Alamos National Laboratory\\
        E-mail: \email{boram@lanl.gov}}

\abstract{The quark chromo-electric dipole moment operator and the
  pseudo-scalar fermion bilinear with which it mixes under
  renormalization can both be included in a calculation of the
  electromagnetic form factor of the nucleon using the Schwinger
  source method. A preliminary calculation of these operators using
  clover quarks on HISQ lattices generated by MILC collaboration will
  be presented showing the quality of the signal in the correlators
  necessary for calculating the neutron electric dipole moment.}

\FullConference{34th annual International Symposium on Lattice Field Theory\\
		24-30 July 2016\\
		University of Southampton, UK}

\makeatletter
\let\begdoctoks\@begindocumenthook
\def\@begindocumenthook{}
\makeatother
\usepackage[colorlinks=true,hyperfootnotes=false]{hyperref}
\hypersetup{allcolors=[cmyk]{0.9,0.9,0,0}}
\AtBeginDocument{\begdoctoks}
\begin{document}

\section{Introduction}

The standard model (SM) of particle physics is almost symmetric under the combined operation of charge-conjugation and parity (CP). In the hadronic sector, the violations come from three sources: a possible term proportional to QCD instanton density, possible CP-violating mass matrix of the quarks, and a complex phase in the Cabibo-Kobayashi-Maskawa (CKM) quark mixing matrix describing the weak interactions.  The axial U(1) anomaly makes the first two of these physically equivalent, and the combined effect, usually parameterized by a quantity called \(\bar\theta\), is experimentally constrained to an unnaturally small value.  The physical effect of the CKM phases is also small, since it is suppressed by the quark masses.

On the other hand, the cosmological abundance of baryons over anti-baryons needs a larger CP-violation.  If this violation is in the hadronic physics, it could lead to an experimentally observable neutron electric dipole moment (nEDM). In fact, upcoming nEDM searches have the potential to constrain many theories of physics beyond the SM (BSM). Most of these theories modify the standard model at a high energy scale, \(M_{\rm BSM}\). A fruitful way of constraining these theories is to parameterize their low energy effects in terms of an effective field theory expansion in terms of operators whose contributions are suppressed by increasing inverse powers of this high scale.

The leading CP-violating effects in this expansion are encoded in   \(\bar\theta\), discussed above. 
Beyond this, the dimension 5 electric dipole moment (qEDM) and the chromo-electric dipole moments (qCEDM) of fermions are suppressed by \(v_{\rm EW}/M_{\rm BSM}^2\), where \(v_{\rm EW}\) is the scale of the weak interactions.  In many BSM theories, the coefficients of these terms have the same origin as the Yukawa couplings in the standard model, and hence are unnaturally small.  In such cases, their contribution might be comparable to some of the dimension 6 operators suppressed only by \(1/M_{\rm BSM}^2\), which include the Weinberg `gluon chromo-electric' operator and various four-Fermi operators.  Nevertheless, because of their lower dimension, the effects of the qEDM and qCEDM should, in any case, be studied separately.  In previous work~\cite{Yoon:2016jzj,Bhattacharya:2016zcn,Bhattacharya:2015esa}, we have extensively studied nEDM from the qEDM.  In this work, we describe preliminary studies of nEDM from qCEDM.

\section{Lattice Methodology}
nEDM can be extracted by expanding the matrix element of the vector current \(V_\mu\) in the neutron state in terms of the Lorentz covariant form-factors, \(F_{1,2,3,A}\):
\begin{equation}
% \langle N | V_\mu(q) | N \rangle & = &
   \overline {u}_N \left[ 
         \gamma_\mu\;F_1(q^2) + i \frac{[\gamma_\mu,\gamma_\nu]}2 q_\nu\; \frac{F_2(q^2)}{2 m_N} + {(2 i\,m_N \gamma_5 q_\mu - \gamma_\mu \gamma_5 q^2)\;\frac{F_A(q^2)}{m_N^2}} + {\frac{[\gamma_\mu, \gamma_\nu]}2 q_\nu \gamma_5\;\frac{F_3(q^2)}{2 m_N}} \right] u_N\,,
\end{equation}
where we have used the Euclidean notation (\(\gamma_\mu^2 = 1\), \(q_4 = i(m-E)\)) and nEDM is given by \(F_3/2 m_N\).  As described previously~\cite{Bhattacharya:2016oqm}, the lattice calculation of the neutron matrix element in the presence of a qCEDM can be described diagrammatically as
\begin{equation}
\vcenter{\hbox{\includegraphics[width=0.12\textwidth]{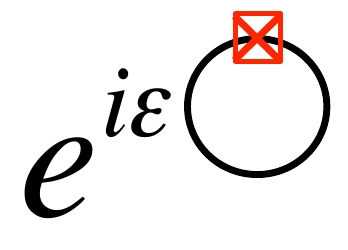}}} \times 
\left(\;\vcenter{\hbox{\includegraphics[width=0.3\textwidth]{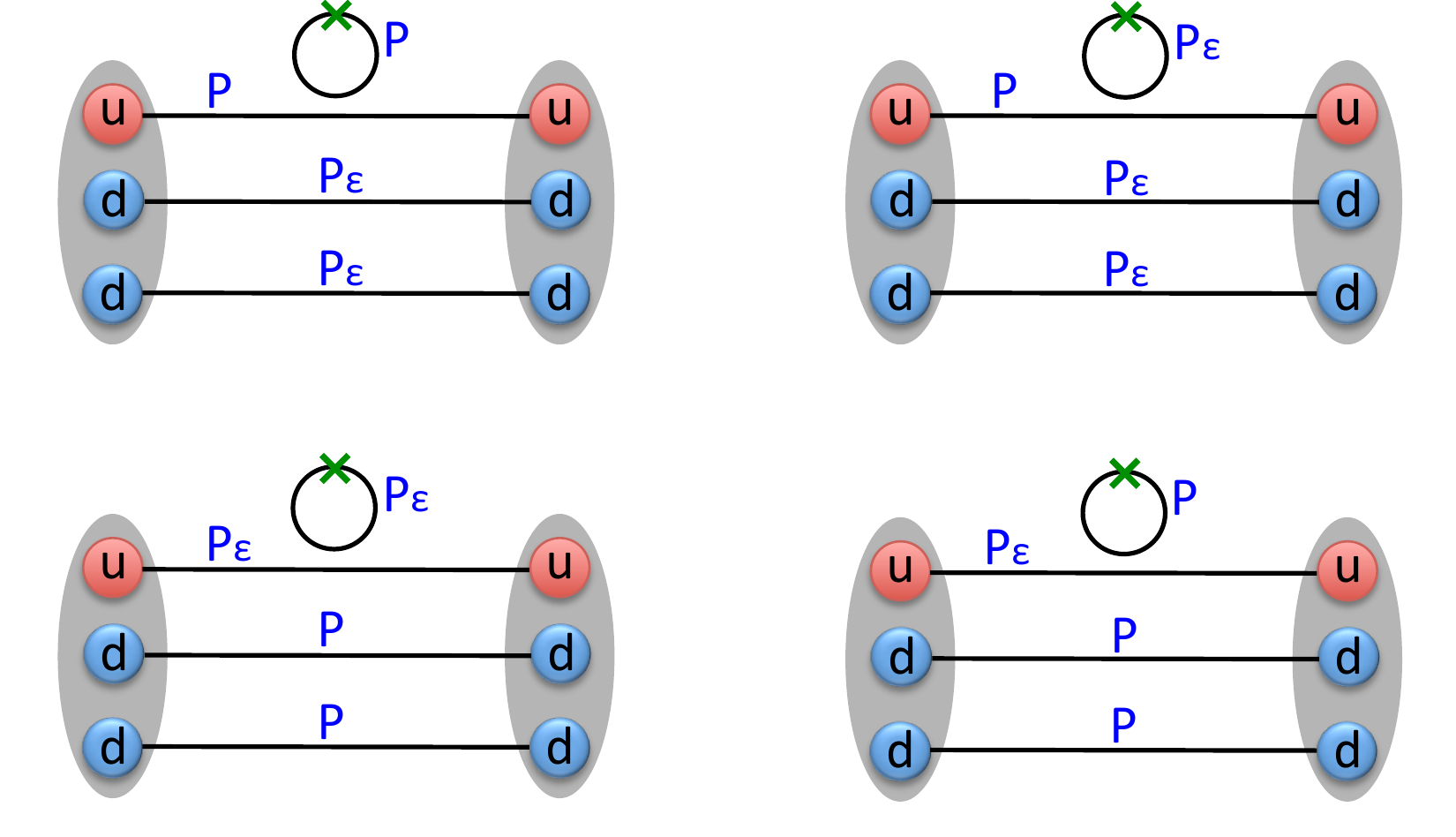}}} +
 \vcenter{\hbox{\includegraphics[width=0.3\textwidth]{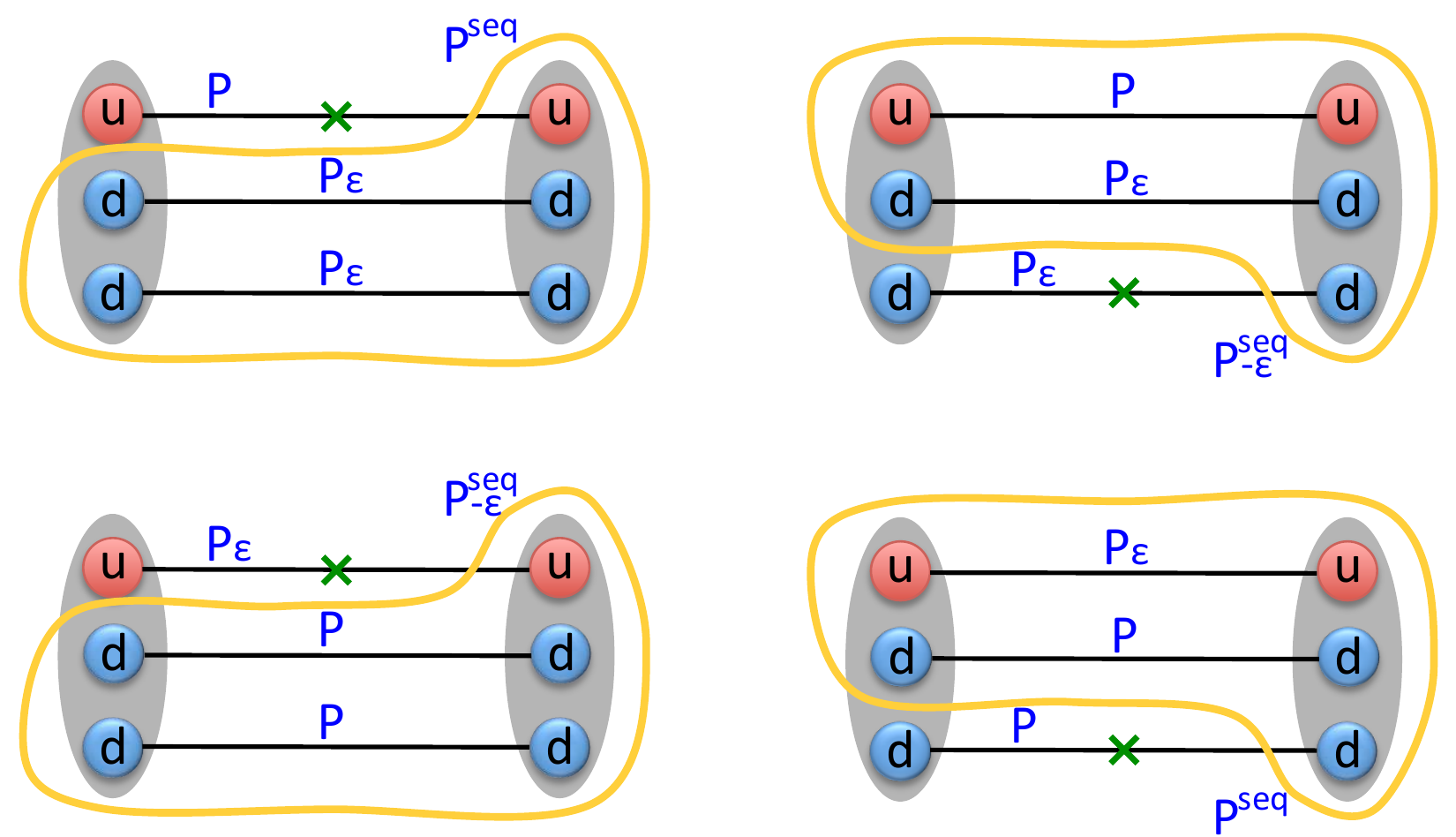}}}\;\right)+O(\epsilon^2)\,.
\end{equation}
Here the two kinds of propagators \(P\) and \(P_\epsilon\) refer to the inverse of the Dirac operator describing the propagation of quarks in a gauge background without and with a CP-violating operator (qCEDM or the CP-violating mass term, \(\bar\psi\gamma_5\psi\)) with coefficient \(\epsilon\equiv\epsilon_5\).  It is to be noted that the \(O(\epsilon^2)\) terms are ultraviolet divergent when propagators include operators of dimension greater than 4, so one should check that \(\epsilon\) is chosen small enough to stay in the linear regime in the ultraviolet regulated theory.\looseness-1

In this preliminary work, we neglect all the diagrams with disconnected contributions including the reweighting exponential factor.  We use two ensembles with lattice spacings \(a\approx0.12\) and \(0.09~{\rm fm}\) in this work: both generated with the 2+1+1 flavors of HISQ~\cite{Follana:2006rc} quarks and with the pion mass, \(M_\pi \approx 310~{\rm MeV}\) by the MILC collaboration~\cite{Bazavov:2012xda}.  We use clover~\cite{Sheikholeslami:1985ij} valence quarks, with Wilson parameter \(\kappa=0.1272103\) and \(0.1266265\) and clover coefficient \(c_{\rm SW}=1.05094\) and \(1.04243\), respectively, on the two ensembles, and use the AMA variance reduction techniques~\cite{Bali:2009hu,Blum:2012uh} with 64 low-precision (relative residual error \(10^{-3}\)) and 4 high-precision (residual \(10^{-8}\)) inversions on each of 400 and 270 configurations in the two ensembles respectively.  In each case, we calculate the matrix elements with both the qCEDM and the CP-violating pseudoscalar mass term since these operators mix under renormalization.

\section{Two point function}

\begin{figure}
  \begin{center}
    \includegraphics[width=0.9\textwidth]{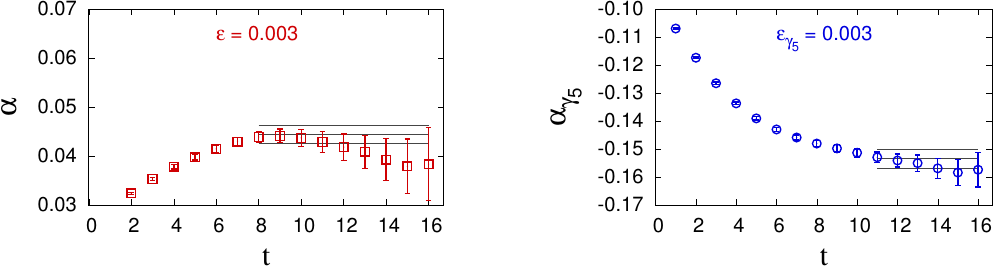}\\
    \includegraphics[width=0.9\textwidth]{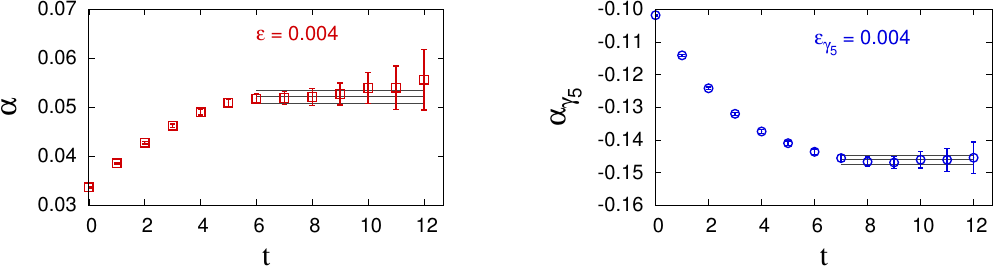}
  \end{center}
  \caption{Connected \(\alpha\) for the \(a\approx0.09~\rm fm\) (top) and \(0.12~\rm fm\) (bottom) ensembles with \(\epsilon=0.003\) and \(0.004\) respectively.
    The left figures show the \(\alpha\) due to the qCEDM operator and the right ones give the analogous quantity \(\alpha_{\gamma_5}\) due to the CP-violating 
    mass term. \label{fig:alpha}}
\end{figure}

We construct the nucleon operator, \(\chi\), out of covariant Gaussian-smeared~\cite{Alexandrou:1990dq} quark fields, \(q_f^c\) as
\(\chi \equiv \epsilon^{abd} [q_1^{aT} C \gamma_5 \frac12(1\pm\gamma_4)q_2^b]q_1^d\), where \(f\) and \(c\) denote flavor and color labels and \(C\) denotes the charge conjugation matrix, and the sign is chosen to be positive and negative for forward and backward propagation. We chose a smearing radius 5.0 and 6.5 lattice units on the \(a\approx 0.12~\rm fm\) and \(0.09~\rm fm\) lattices (with 46 and 85 Gaussian hits) respectively, so that the physical radius is about \(0.6~\rm fm\). Because of the spectral representation and Lorentz invariance, the nucleon propagator \(\langle0|\chi(\vec p, t) \bar\chi(\vec p, 0)|0\rangle\) approaches \(e^{-p_4 t} e^{i\alpha\gamma_5} (\slashed p + m) e^{i\alpha\gamma_5}\) as \(t\to\infty\), where \(\alpha=0\) unless CP is violated. We can use the measured two point function to provide an estimate of \(\alpha\), the phase in the neutron wavefunction, as a function of the CP-violation parameter in the action.  In Fig.~\ref{fig:alpha}, we show the signal in the extraction of this quantity from the neutron two-point correlator.

\begin{figure}
  \begin{center}
    \colorbox{white}{%
      \includegraphics[width=0.45\textwidth]{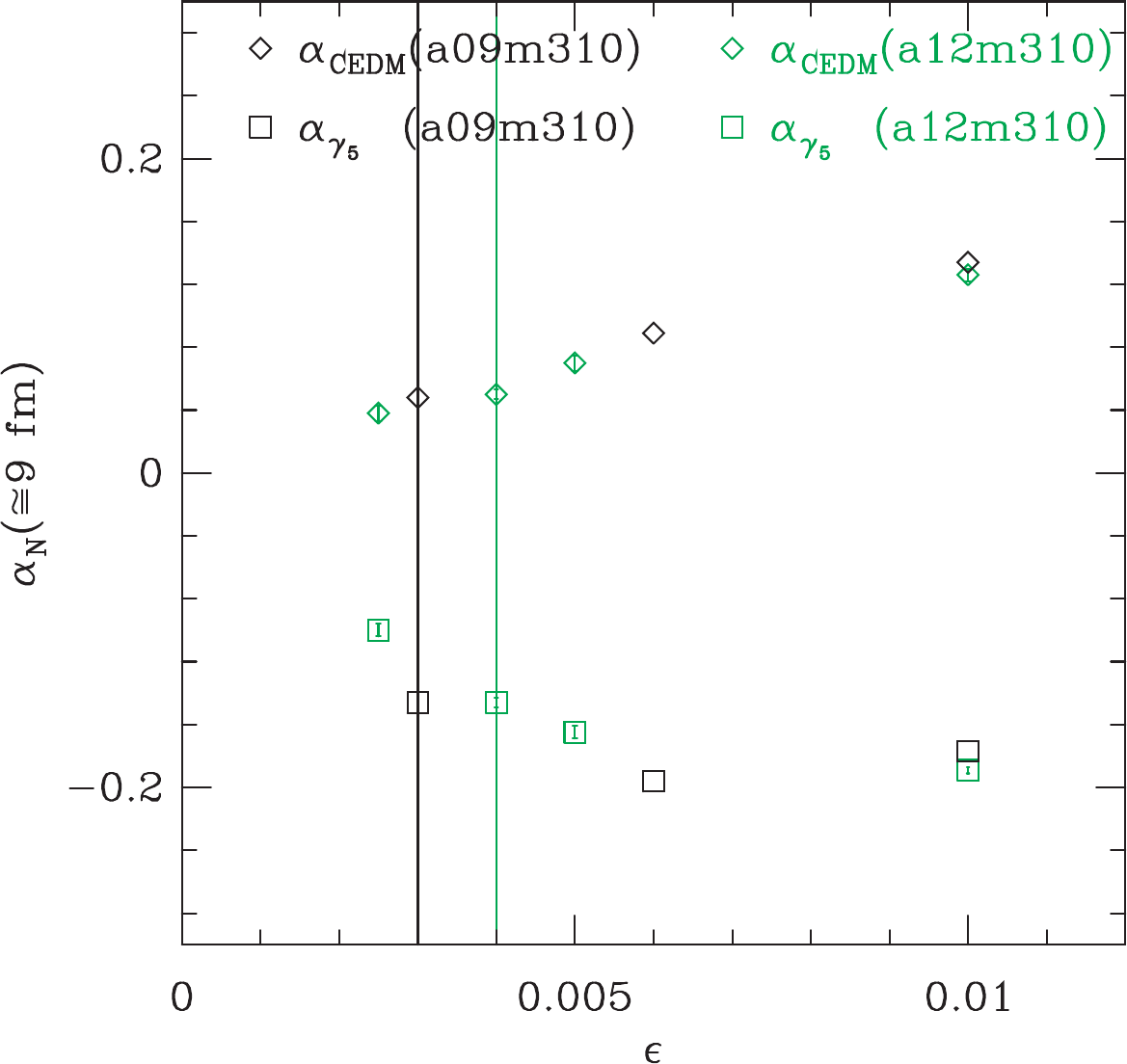}%
    }
  \end{center}
  \caption{Linearity of connected \(\alpha\) with \(\epsilon\).}
  \label{fig:linearity}
\end{figure}

In Fig.~\ref{fig:linearity}, we show that the measured value of \(\alpha\) is roughly linear in \(\epsilon\) around \(\epsilon\sim0.005\) for both the qCEDM and the CP-violating quark mass operators.  Based on this, we choose \(\epsilon\approx (30 fm)^{-1}a \approx 6.6 {\rm MeV}a \approx 0.36 ma\), where \(m\) is the quark mass for further analysis.

\begin{figure}
  \begin{center}
    \includegraphics[width=0.45\textwidth]{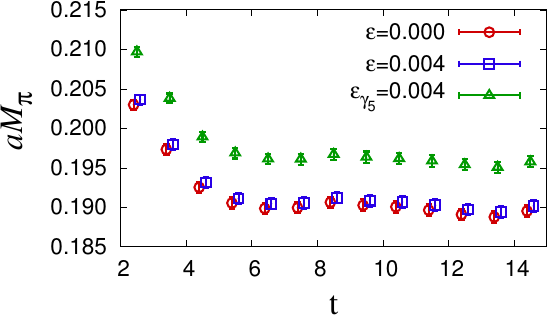}
    \hspace{0.05\textwidth}
    \includegraphics[width=0.45\textwidth]{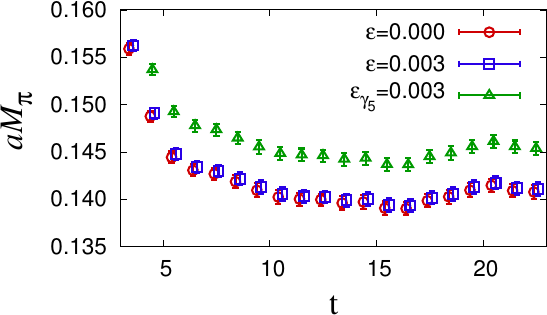}
  \end{center}
  \caption{Pion effective mass for \(a\approx0.12~\rm fm\) (left) and \(a\approx0.09~\rm fm\) (right) ensembles.
    \label{fig:Mpi}}
\end{figure}

Since the chiral and CP rotations do not commute, one can use a chiral rotation of the fermion fields to remove the CP-violation in the mass term~\cite{Bhattacharya:2015rsa}. Such a rotation changes the mass term \((m a)^2 \to (m a)^2 + \epsilon^2\) and introduces an \(O(a)\) CP-violation from the non-invariant Wilson and clover terms in the action that we discuss later.  In Figure~\ref{fig:Mpi} we show the effective mass plots for the pion propagator: as expected, addition of the CP-violating mass term affects the pion mass much more than the addition of the higher dimensional qCEDM operator.  Quantitatively, in the lowest order chiral perturbation theory ($\chi$PT), one would expect the pion mass to increase by a factor of \(\sqrt{1+\frac{\epsilon^2}{(ma)^2}}\). In Tab.~\ref{tab:Mpi}, we show that this expectation is indeed consistent with the measurement.

\begin{table}
\begin{center}
\begin{tabular}{|c|r|r|}
\hline
&a12m310&a09m310\\
\hline
\(\frac1{2\kappa}-4\)&-0.0695&-0.05138\\
\(\frac1{2\kappa_c}-4\)&-0.08058&-0.05943\\
\(a m \equiv \frac1{2\kappa}-\frac1{2\kappa_c}\)&0.01108&0.00859\\
\(\epsilon\)&0.004&0.003\\
\hline
\(a M_\pi^0\)&0.1900(4)&0.1404(3)\\
\(a M_\pi^{\rm qCEDM}\)&0.1906(4)&0.1407(3)\\
\(a M_\pi^{\rm P}\)&0.1961(4)&0.1450(3)\\
\(a M_\pi^0 \sqrt{1+\frac{\epsilon^2}{(ma)^2}}\)&0.1959(4)&0.1450(3)\\
\hline
\end{tabular}
\end{center}
\caption{The pion mass in the theory without CP-violation, \(M_\pi^0\), in the theory with qCEDM operator, \(M_\pi^{\rm qCEDM}\), and in the theory with CP-violating mass term \(M_\pi^{\rm P}\).  Also provided are the quark mass, \(a m\), and the coefficient, \(\epsilon\), of the CP-violating operator. \(M_\pi^{\rm P}\) is compared to its lowest order $\chi$PT expectation given in the last row.}
\label{tab:Mpi}
\end{table}

\section{Three-point functions}

\begin{figure}
  \begin{center}
      \includegraphics[width=0.2\textwidth]{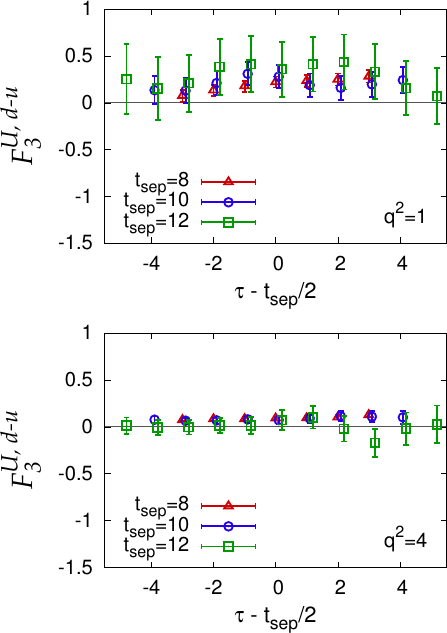}%
      \hspace{0.05\textwidth}%
      \includegraphics[width=0.2\textwidth]{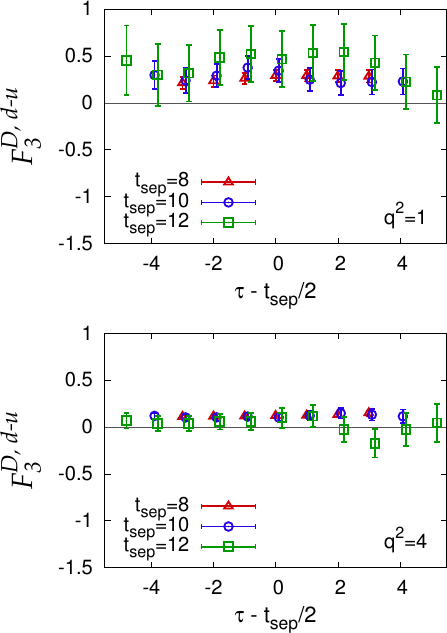}%
      \hspace{0.05\textwidth}%
      \includegraphics[width=0.2\textwidth]{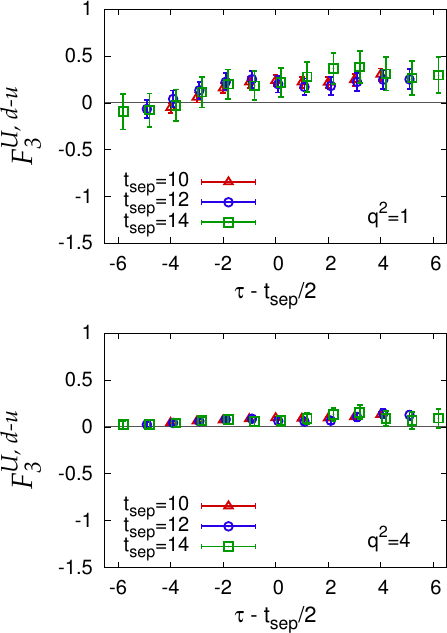}%
      \hspace{0.05\textwidth}%
      \includegraphics[width=0.2\textwidth]{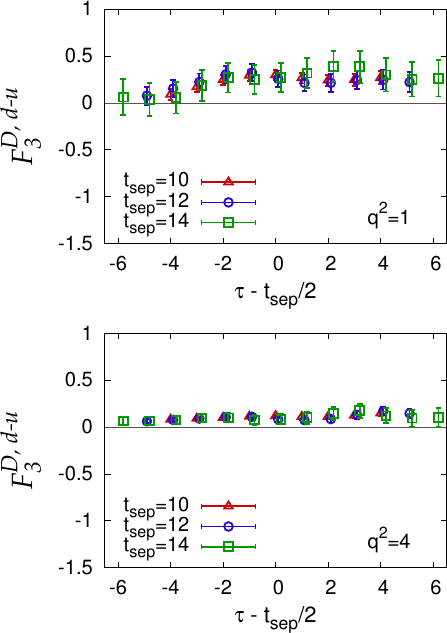}

      \includegraphics[width=0.2\textwidth]{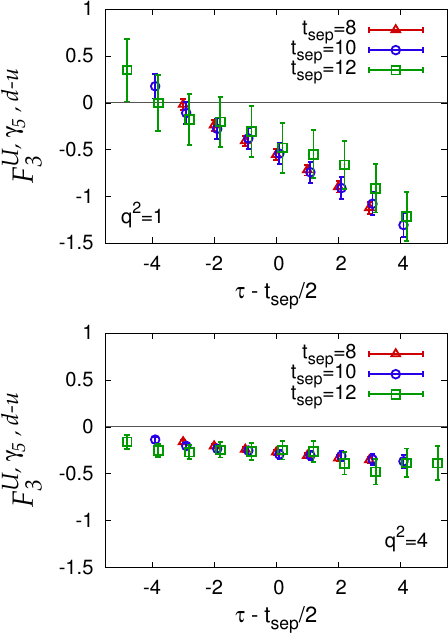}%
      \hspace{0.05\textwidth}%
      \includegraphics[width=0.2\textwidth]{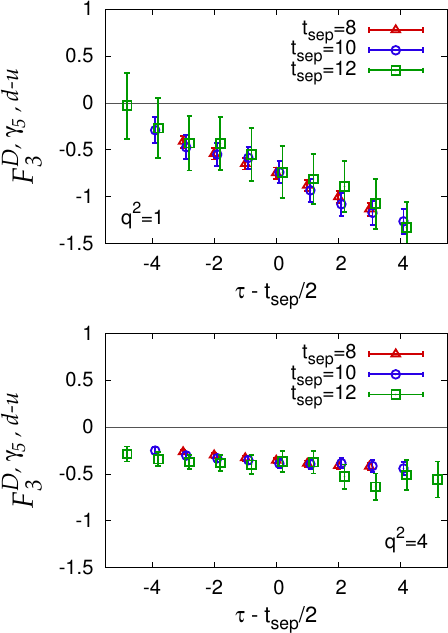}%
      \hspace{0.05\textwidth}%
      \includegraphics[width=0.2\textwidth]{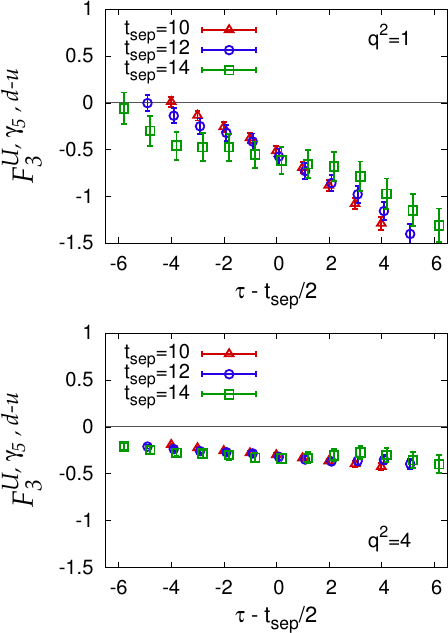}%
      \hspace{0.05\textwidth}%
      \includegraphics[width=0.2\textwidth]{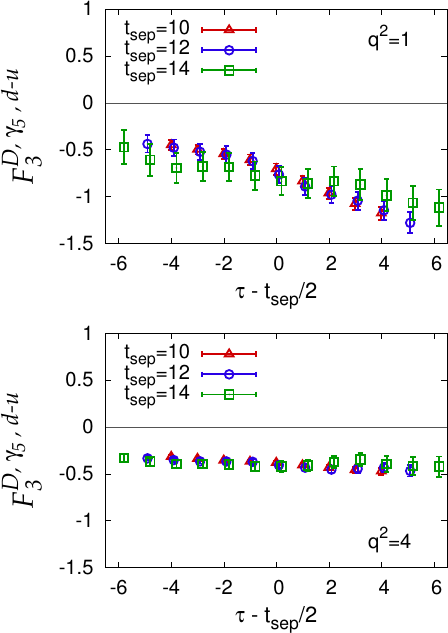}
  \end{center}
  \caption{Connected \(F_3\) from qCEDM (top two rows) and CP-violating mass (bottom two rows) measured on \(a\approx0.12~\rm fm\) (left two columns) and \(0.09~\rm fm\) (right two columns) ensembles.\label{fig:F3}}
\end{figure}

The neutron three-point function of the vector-current component, \(\langle V_3(q)\rangle \equiv \langle 0 | N V_3(q) \bar N | 0 \rangle\), projected with \({\cal P}\equiv(1+\gamma_4)(1+i\gamma_5\gamma_3)/2\) can be calculated with the fermion source having a single non-zero spin component. Straightforward calculation reveals that with this projection the Lorentz-invariant form factors are
% (here $G_E = F_1 - q^2/(2 m_N) F_2$):
\begin{eqnarray}
  \Tr {\cal P}\langle V_3(q) \rangle &\propto& i m_N q_3 
  %G_E
  [F_1 - \frac{q^2}{2 m_N} F_2 ] 
  + \alpha [ m_N (E_N-m_N) (F_1 + F_2) + \frac{q_3^2}2F_2]
  \nonumber \\
   & -& 2i (q_1^2+q_2^2) F_A - \frac{q_3^2}2F_3\,,
\end{eqnarray}
which, along with the determination of \(\alpha\) from the two-point function, can be used to extract \(F_3\). In Fig.~\ref{fig:F3}, we show the signal in the determination of the connected contribution to the isovector \(F_3^{q,d-u}\), \(q=U,D\), from various source-sink separations for CP-violation, in the up and down quark sectors respectively, arising due to the qCEDM or parity-violating mass terms.  In each case, the calculation is shown for two values of momenta, \(\vec q=(0,0,1)\) and \((0,0,2)\).  A plateau is visible in all these calculations, especially at large source-sink separation, except for the coarse ensemble with CP-violating mass term and one unit of momentum.

\begin{figure}
  \begin{center}
  \includegraphics[width=0.45\textwidth]{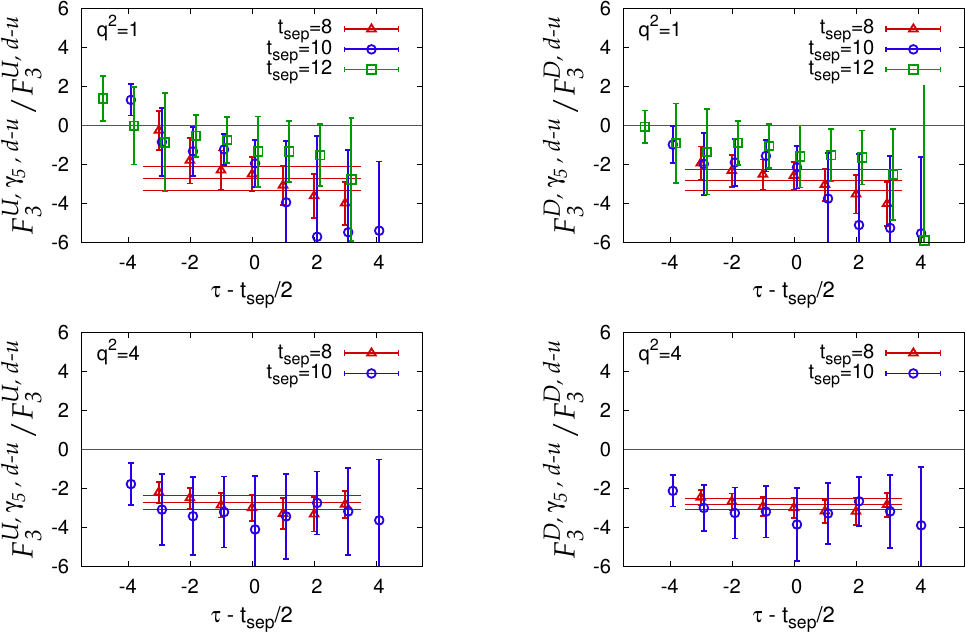}\hspace{0.04\textwidth}
  \includegraphics[width=0.45\textwidth]{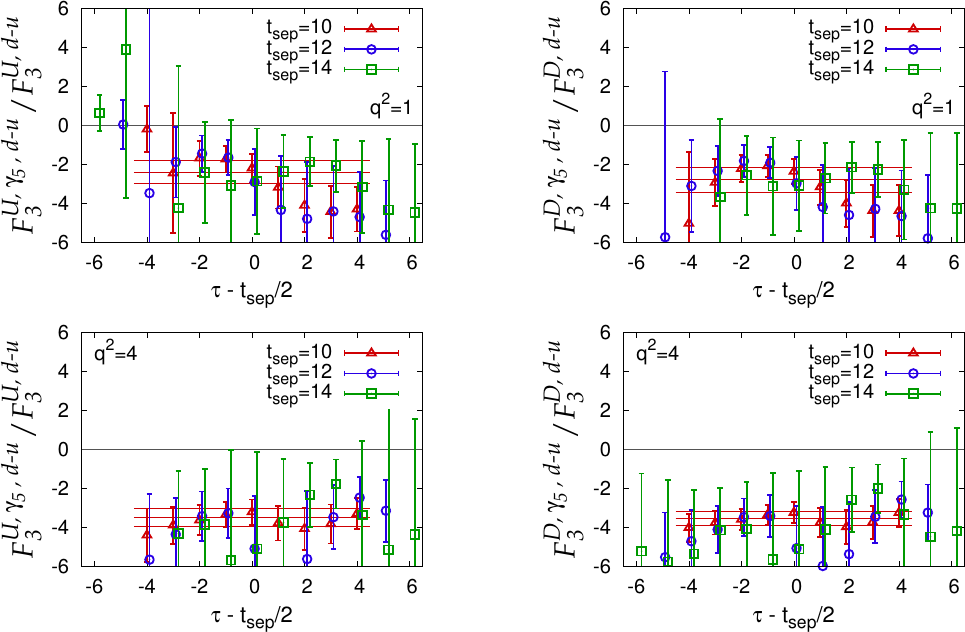}
  \end{center}
  \caption{Ratio of F3 from CP-violating mass to that from qCEDM from \(a\approx0.12~\rm fm\) (left two columns) and \(0.09~\rm fm\) (right two columns) ensembles. \label{fig:rat}}
\end{figure}

It is, however, known that a CP-violating mass term should result in zero contribution to \(F_3\) from connected diagrams~\cite{Aoki:1990ix,Guadagnoli:2002nm}, since such a mass term can be rotated away by a chiral rotation.  That argument, however, ignores \(O(a)\) effects.  In fact, at tree-level, we can write
\begin{equation}
  \bar T(\alpha_\epsilon,\zeta_\epsilon) \left[\Dslash_W(r,\chi) + me^{i\frac\epsilon m\gamma_5}\right] T(\alpha_\epsilon,\zeta_\epsilon) = z_\epsilon\Dslash_W\left(\frac{r_\epsilon}{z_\epsilon},
  \frac{\chi_\epsilon}{z_\epsilon}\right) + m_\epsilon + ia\tilde\xi_\epsilon \Sigma^{\mu\nu}G_{\mu\nu}\gamma_5 + O(a^2)\,,
\end{equation}
where
\begin{eqnarray}
&& \Dslash_W(r,\chi) \equiv \Dslash + a (r D^2 + \chi \Sigma^{\mu\nu}G_{\mu\nu})\,,\quad
  T(\alpha,\zeta) \equiv e^{i\frac\zeta2a[\Dslash_W(r,\chi)-me^{i\frac\epsilon m\gamma_5}]\gamma_5}e^{i\frac\alpha2\gamma_5}\,,\nonumber\\
&&  \bar T(\alpha,\zeta)\equiv\gamma_0T^\dagger(\alpha,\zeta)\gamma_0\,,\quad
  \zeta_\epsilon \equiv r \tan\frac\epsilon m \frac1{1+ram\cos\frac\epsilon m}\,,\quad \alpha_\epsilon \equiv -\tan^{-1}\frac{\zeta_\epsilon}r\,,\nonumber\\
  &&z_\epsilon \equiv 1 + 2am\zeta_\epsilon\sin\frac\epsilon m\,,\quad\xi_\epsilon \equiv -\frac\epsilon m + \tan^{-1}\left(\frac {gr}{2\chi} \tan \frac \epsilon m\right)\,,\quad \tilde\xi_\epsilon \equiv \frac{\chi_\epsilon}{z_\epsilon}\sin\xi_\epsilon\,,\nonumber\\
  && m_\epsilon \equiv m\sqrt{1+2am\zeta_\epsilon\sin\frac\epsilon m+\zeta_\epsilon^2(am)^2}\,,\quad r_\epsilon \equiv r \sec \frac\epsilon m \,,\quad\chi_\epsilon \equiv \chi\sec\left(\xi_\epsilon+\frac\epsilon m\right)\,,
\end{eqnarray}
and \(g\) is the strong coupling constant. Because of this one expects that the on-shell connected vector-current matrix element in a Wilson-clover theory with CP-violating mass term should be 
% almost
 proportional\footnote{{up to \(O(a^2)\).}}
 to the same matrix element  in a theory with CP-violation arising from the qCEDM operator and a slightly shifted mass \(m\), Wilson parameter \(r\), and clover coefficient \(\chi\).
  Ignoring these small  $O(\epsilon^2)$ shifts in the fermion parameters, one would then expect the ratio of the \(F_3\) calculated with the CP-violating mass term and with the qCEDM operators to be proportional.  In Fig.~\ref{fig:rat}, we show that this expectation is actually satisfied by the lattice data.

\section{Conclusion}

We have shown reasonable signal-to-noise ratio for calculating the connected \(F_3\) from qCEDM operator on our ensembles.  The significant ultraviolet divergent mixing of this operator is with the CP-violating mass term: we showed that the latter is almost proportional to the contribution of the qCEDM operator itself; so, for these connected contributions, the power divergence is essentially multiplicative. The connected contribution of the CP-violating mass term to \(F_3\) vanishes in formulations with exact chiral symmetry, but the proportionality observed here is expected to hold for {\em any} residual chiral violation. As a result, the connected contribution to \(F_3\) from qCEDM has only multiplicative power divergences in any formulation.

\section{Acknowledgments}
We thank the MILC Collaboration~\cite{Bazavov:2012xda} for providing us with the ensembles. The work was supported by the U.S. DoE HEP Office of Science contract number DE-KA-1401020 and the LANL LDRD Program.   All simulations were carried out using IC resources at LANL.

\bibliographystyle{plain}
\bibliography{skeleton}

\end{document}